\pdfoutput=1
\documentclass[11pt,twoside,a4paper,pdftex,cmspaper,final,collab]{cms-tdr}

\begin{document}\cmsNoteHeader{CFT-09-006}
%
%
%

%
%
\hyphenation{env-iron-men-tal}
\hyphenation{had-ron-i-za-tion}
\hyphenation{cal-or-i-me-ter}
\hyphenation{de-vices}
%
%
\RCS$Revision: 1.27 $
\RCS$Date: 2009/12/21 12:14:44 $
\RCS$Name:  $
%
%
%

\newcommand {\etal}{\mbox{et al.}\xspace} 
\newcommand {\ie}{\mbox{i.e.}\xspace}     
\newcommand {\eg}{\mbox{e.g.}\xspace}     
\newcommand {\etc}{\mbox{etc.}\xspace}     
\newcommand {\vs}{\mbox{\sl vs.}\xspace}      
\newcommand {\mdash}{\ensuremath{\mathrm{-}}} 

\newcommand {\Lone}{Level-1\xspace} 
\newcommand {\Ltwo}{Level-2\xspace}
\newcommand {\Lthree}{Level-3\xspace}

\providecommand{\ACERMC} {\textsc{AcerMC}\xspace}
\providecommand{\ALPGEN} {{\textsc{alpgen}}\xspace}
\providecommand{\CHARYBDIS} {{\textsc{charybdis}}\xspace}
\providecommand{\CMKIN} {\textsc{cmkin}\xspace}
\providecommand{\CMSIM} {{\textsc{cmsim}}\xspace}
\providecommand{\CMSSW} {{\textsc{cmssw}}\xspace}
\providecommand{\COBRA} {{\textsc{cobra}}\xspace}
\providecommand{\COCOA} {{\textsc{cocoa}}\xspace}
\providecommand{\COMPHEP} {\textsc{CompHEP}\xspace}
\providecommand{\EVTGEN} {{\textsc{evtgen}}\xspace}
\providecommand{\FAMOS} {{\textsc{famos}}\xspace}
\providecommand{\GARCON} {\textsc{garcon}\xspace}
\providecommand{\GARFIELD} {{\textsc{garfield}}\xspace}
\providecommand{\GEANE} {{\textsc{geane}}\xspace}
\providecommand{\GEANTfour} {{\textsc{geant4}}\xspace}
\providecommand{\GEANTthree} {{\textsc{geant3}}\xspace}
\providecommand{\GEANT} {{\textsc{geant}}\xspace}
\providecommand{\HDECAY} {\textsc{hdecay}\xspace}
\providecommand{\HERWIG} {{\textsc{herwig}}\xspace}
\providecommand{\HIGLU} {{\textsc{higlu}}\xspace}
\providecommand{\HIJING} {{\textsc{hijing}}\xspace}
\providecommand{\IGUANA} {\textsc{iguana}\xspace}
\providecommand{\ISAJET} {{\textsc{isajet}}\xspace}
\providecommand{\ISAPYTHIA} {{\textsc{isapythia}}\xspace}
\providecommand{\ISASUGRA} {{\textsc{isasugra}}\xspace}
\providecommand{\ISASUSY} {{\textsc{isasusy}}\xspace}
\providecommand{\ISAWIG} {{\textsc{isawig}}\xspace}
\providecommand{\MADGRAPH} {\textsc{MadGraph}\xspace}
\providecommand{\MCATNLO} {\textsc{mc@nlo}\xspace}
\providecommand{\MCFM} {\textsc{mcfm}\xspace}
\providecommand{\MILLEPEDE} {{\textsc{millepede}}\xspace}
\providecommand{\ORCA} {{\textsc{orca}}\xspace}
\providecommand{\OSCAR} {{\textsc{oscar}}\xspace}
\providecommand{\PHOTOS} {\textsc{photos}\xspace}
\providecommand{\PROSPINO} {\textsc{prospino}\xspace}
\providecommand{\PYTHIA} {{\textsc{pythia}}\xspace}
\providecommand{\SHERPA} {{\textsc{sherpa}}\xspace}
\providecommand{\TAUOLA} {\textsc{tauola}\xspace}
\providecommand{\TOPREX} {\textsc{TopReX}\xspace}
\providecommand{\XDAQ} {{\textsc{xdaq}}\xspace}

\newcommand {\DZERO}{D\O\xspace}     


\newcommand{\de}{\ensuremath{^\circ}}
\newcommand{\ten}[1]{\ensuremath{\times \text{10}^\text{#1}}}
\newcommand{\unit}[1]{\ensuremath{\text{\,#1}}\xspace}
\newcommand{\mum}{\ensuremath{\,\mu\text{m}}\xspace}
\newcommand{\micron}{\ensuremath{\,\mu\text{m}}\xspace}
\newcommand{\cm}{\ensuremath{\,\text{cm}}\xspace}
\newcommand{\mm}{\ensuremath{\,\text{mm}}\xspace}
\newcommand{\mus}{\ensuremath{\,\mu\text{s}}\xspace}
\newcommand{\keV}{\ensuremath{\,\text{ke\hspace{-.08em}V}}\xspace}
\newcommand{\MeV}{\ensuremath{\,\text{Me\hspace{-.08em}V}}\xspace}
\newcommand{\GeV}{\ensuremath{\,\text{Ge\hspace{-.08em}V}}\xspace}
\newcommand{\TeV}{\ensuremath{\,\text{Te\hspace{-.08em}V}}\xspace}
\newcommand{\PeV}{\ensuremath{\,\text{Pe\hspace{-.08em}V}}\xspace}
\newcommand{\keVc}{\ensuremath{{\,\text{ke\hspace{-.08em}V\hspace{-0.16em}/\hspace{-0.08em}c}}}\xspace}
\newcommand{\MeVc}{\ensuremath{{\,\text{Me\hspace{-.08em}V\hspace{-0.16em}/\hspace{-0.08em}c}}}\xspace}
\newcommand{\GeVc}{\ensuremath{{\,\text{Ge\hspace{-.08em}V\hspace{-0.16em}/\hspace{-0.08em}c}}}\xspace}
\newcommand{\TeVc}{\ensuremath{{\,\text{Te\hspace{-.08em}V\hspace{-0.16em}/\hspace{-0.08em}c}}}\xspace}
\newcommand{\keVcc}{\ensuremath{{\,\text{ke\hspace{-.08em}V\hspace{-0.16em}/\hspace{-0.08em}c}^\text{2}}}\xspace}
\newcommand{\MeVcc}{\ensuremath{{\,\text{Me\hspace{-.08em}V\hspace{-0.16em}/\hspace{-0.08em}c}^\text{2}}}\xspace}
\newcommand{\GeVcc}{\ensuremath{{\,\text{Ge\hspace{-.08em}V\hspace{-0.16em}/\hspace{-0.08em}c}^\text{2}}}\xspace}
\newcommand{\TeVcc}{\ensuremath{{\,\text{Te\hspace{-.08em}V\hspace{-0.16em}/\hspace{-0.08em}c}^\text{2}}}\xspace}

\newcommand{\pbinv} {\mbox{\ensuremath{\,\text{pb}^\text{$-$1}}}\xspace}
\newcommand{\fbinv} {\mbox{\ensuremath{\,\text{fb}^\text{$-$1}}}\xspace}
\newcommand{\nbinv} {\mbox{\ensuremath{\,\text{nb}^\text{$-$1}}}\xspace}
\newcommand{\percms}{\ensuremath{\,\text{cm}^\text{$-$2}\,\text{s}^\text{$-$1}}\xspace}
\newcommand{\lumi}{\ensuremath{\mathcal{L}}\xspace}
\newcommand{\Lumi}{\ensuremath{\mathcal{L}}\xspace}
%
\newcommand{\LvLow}  {\ensuremath{\mathcal{L}=\text{10}^\text{32}\,\text{cm}^\text{$-$2}\,\text{s}^\text{$-$1}}\xspace}
\newcommand{\LLow}   {\ensuremath{\mathcal{L}=\text{10}^\text{33}\,\text{cm}^\text{$-$2}\,\text{s}^\text{$-$1}}\xspace}
\newcommand{\lowlumi}{\ensuremath{\mathcal{L}=\text{2}\times \text{10}^\text{33}\,\text{cm}^\text{$-$2}\,\text{s}^\text{$-$1}}\xspace}
\newcommand{\LMed}   {\ensuremath{\mathcal{L}=\text{2}\times \text{10}^\text{33}\,\text{cm}^\text{$-$2}\,\text{s}^\text{$-$1}}\xspace}
\newcommand{\LHigh}  {\ensuremath{\mathcal{L}=\text{10}^\text{34}\,\text{cm}^\text{$-$2}\,\text{s}^\text{$-$1}}\xspace}
\newcommand{\hilumi} {\ensuremath{\mathcal{L}=\text{10}^\text{34}\,\text{cm}^\text{$-$2}\,\text{s}^\text{$-$1}}\xspace}


\newcommand{\zp}{\ensuremath{\mathrm{Z}^\prime}\xspace}


\newcommand{\kt}{\ensuremath{k_{\mathrm{T}}}\xspace}
\newcommand{\BC}{\ensuremath{{B_{\mathrm{c}}}}\xspace}
\newcommand{\bbarc}{\ensuremath{{\overline{b}c}}\xspace}
\newcommand{\bbbar}{\ensuremath{{b\overline{b}}}\xspace}
\newcommand{\ccbar}{\ensuremath{{c\overline{c}}}\xspace}
\newcommand{\JPsi}{\ensuremath{{J}/\psi}\xspace}
\newcommand{\bspsiphi}{\ensuremath{B_s \to \JPsi\, \phi}\xspace}
\newcommand{\AFB}{\ensuremath{A_\mathrm{FB}}\xspace}
\newcommand{\EE}{\ensuremath{e^+e^-}\xspace}
\newcommand{\MM}{\ensuremath{\mu^+\mu^-}\xspace}
\newcommand{\TT}{\ensuremath{\tau^+\tau^-}\xspace}
\newcommand{\wangle}{\ensuremath{\sin^{2}\theta_{\mathrm{eff}}^\mathrm{lept}(M^2_\mathrm{Z})}\xspace}
\newcommand{\ttbar}{\ensuremath{{t\overline{t}}}\xspace}
\newcommand{\stat}{\ensuremath{\,\text{(stat.)}}\xspace}
\newcommand{\syst}{\ensuremath{\,\text{(syst.)}}\xspace}

\newcommand{\HGG}{\ensuremath{\mathrm{H}\to\gamma\gamma}}
\newcommand{\gev}{\GeV}
\newcommand{\GAMJET}{\ensuremath{\gamma + \mathrm{jet}}}
\newcommand{\PPTOJETS}{\ensuremath{\mathrm{pp}\to\mathrm{jets}}}
\newcommand{\PPTOGG}{\ensuremath{\mathrm{pp}\to\gamma\gamma}}
\newcommand{\PPTOGAMJET}{\ensuremath{\mathrm{pp}\to\gamma +
\mathrm{jet}
}}
\newcommand{\MH}{\ensuremath{\mathrm{M_{\mathrm{H}}}}}
\newcommand{\RNINE}{\ensuremath{\mathrm{R}_\mathrm{9}}}
\newcommand{\DR}{\ensuremath{\Delta\mathrm{R}}}


\newcommand{\PT}{\ensuremath{p_{\mathrm{T}}}\xspace}
\newcommand{\pt}{\ensuremath{p_{\mathrm{T}}}\xspace}
\newcommand{\ET}{\ensuremath{E_{\mathrm{T}}}\xspace}
\newcommand{\HT}{\ensuremath{H_{\mathrm{T}}}\xspace}
\newcommand{\et}{\ensuremath{E_{\mathrm{T}}}\xspace}
\newcommand{\Em}{\ensuremath{E\!\!\!/}\xspace}
\newcommand{\Pm}{\ensuremath{p\!\!\!/}\xspace}
\newcommand{\PTm}{\ensuremath{{p\!\!\!/}_{\mathrm{T}}}\xspace}
\newcommand{\ETm}{\ensuremath{E_{\mathrm{T}}^{\mathrm{miss}}}\xspace}
\newcommand{\MET}{\ensuremath{E_{\mathrm{T}}^{\mathrm{miss}}}\xspace}
\newcommand{\ETmiss}{\ensuremath{E_{\mathrm{T}}^{\mathrm{miss}}}\xspace}
\newcommand{\VEtmiss}{\ensuremath{{\vec E}_{\mathrm{T}}^{\mathrm{miss}}}\xspace}

%

\newcommand{\ga}{\ensuremath{\gtrsim}}
\newcommand{\la}{\ensuremath{\lesssim}}
\newcommand{\swsq}{\ensuremath{\sin^2\theta_W}\xspace}
\newcommand{\cwsq}{\ensuremath{\cos^2\theta_W}\xspace}
\newcommand{\tanb}{\ensuremath{\tan\beta}\xspace}
\newcommand{\tanbsq}{\ensuremath{\tan^{2}\beta}\xspace}
\newcommand{\sidb}{\ensuremath{\sin 2\beta}\xspace}
\newcommand{\alpS}{\ensuremath{\alpha_S}\xspace}
\newcommand{\alpt}{\ensuremath{\tilde{\alpha}}\xspace}

\newcommand{\QL}{\ensuremath{Q_L}\xspace}
\newcommand{\sQ}{\ensuremath{\tilde{Q}}\xspace}
\newcommand{\sQL}{\ensuremath{\tilde{Q}_L}\xspace}
\newcommand{\ULC}{\ensuremath{U_L^C}\xspace}
\newcommand{\sUC}{\ensuremath{\tilde{U}^C}\xspace}
\newcommand{\sULC}{\ensuremath{\tilde{U}_L^C}\xspace}
\newcommand{\DLC}{\ensuremath{D_L^C}\xspace}
\newcommand{\sDC}{\ensuremath{\tilde{D}^C}\xspace}
\newcommand{\sDLC}{\ensuremath{\tilde{D}_L^C}\xspace}
\newcommand{\LL}{\ensuremath{L_L}\xspace}
\newcommand{\sL}{\ensuremath{\tilde{L}}\xspace}
\newcommand{\sLL}{\ensuremath{\tilde{L}_L}\xspace}
\newcommand{\ELC}{\ensuremath{E_L^C}\xspace}
\newcommand{\sEC}{\ensuremath{\tilde{E}^C}\xspace}
\newcommand{\sELC}{\ensuremath{\tilde{E}_L^C}\xspace}
\newcommand{\sEL}{\ensuremath{\tilde{E}_L}\xspace}
\newcommand{\sER}{\ensuremath{\tilde{E}_R}\xspace}
\newcommand{\sFer}{\ensuremath{\tilde{f}}\xspace}
\newcommand{\sQua}{\ensuremath{\tilde{q}}\xspace}
\newcommand{\sUp}{\ensuremath{\tilde{u}}\xspace}
\newcommand{\suL}{\ensuremath{\tilde{u}_L}\xspace}
\newcommand{\suR}{\ensuremath{\tilde{u}_R}\xspace}
\newcommand{\sDw}{\ensuremath{\tilde{d}}\xspace}
\newcommand{\sdL}{\ensuremath{\tilde{d}_L}\xspace}
\newcommand{\sdR}{\ensuremath{\tilde{d}_R}\xspace}
\newcommand{\sTop}{\ensuremath{\tilde{t}}\xspace}
\newcommand{\stL}{\ensuremath{\tilde{t}_L}\xspace}
\newcommand{\stR}{\ensuremath{\tilde{t}_R}\xspace}
\newcommand{\stone}{\ensuremath{\tilde{t}_1}\xspace}
\newcommand{\sttwo}{\ensuremath{\tilde{t}_2}\xspace}
\newcommand{\sBot}{\ensuremath{\tilde{b}}\xspace}
\newcommand{\sbL}{\ensuremath{\tilde{b}_L}\xspace}
\newcommand{\sbR}{\ensuremath{\tilde{b}_R}\xspace}
\newcommand{\sbone}{\ensuremath{\tilde{b}_1}\xspace}
\newcommand{\sbtwo}{\ensuremath{\tilde{b}_2}\xspace}
\newcommand{\sLep}{\ensuremath{\tilde{l}}\xspace}
\newcommand{\sLepC}{\ensuremath{\tilde{l}^C}\xspace}
\newcommand{\sEl}{\ensuremath{\tilde{e}}\xspace}
\newcommand{\sElC}{\ensuremath{\tilde{e}^C}\xspace}
\newcommand{\seL}{\ensuremath{\tilde{e}_L}\xspace}
\newcommand{\seR}{\ensuremath{\tilde{e}_R}\xspace}
\newcommand{\snL}{\ensuremath{\tilde{\nu}_L}\xspace}
\newcommand{\sMu}{\ensuremath{\tilde{\mu}}\xspace}
\newcommand{\sNu}{\ensuremath{\tilde{\nu}}\xspace}
\newcommand{\sTau}{\ensuremath{\tilde{\tau}}\xspace}
\newcommand{\Glu}{\ensuremath{g}\xspace}
\newcommand{\sGlu}{\ensuremath{\tilde{g}}\xspace}
\newcommand{\Wpm}{\ensuremath{W^{\pm}}\xspace}
\newcommand{\sWpm}{\ensuremath{\tilde{W}^{\pm}}\xspace}
\newcommand{\Wz}{\ensuremath{W^{0}}\xspace}
\newcommand{\sWz}{\ensuremath{\tilde{W}^{0}}\xspace}
\newcommand{\sWino}{\ensuremath{\tilde{W}}\xspace}
\newcommand{\Bz}{\ensuremath{B^{0}}\xspace}
\newcommand{\sBz}{\ensuremath{\tilde{B}^{0}}\xspace}
\newcommand{\sBino}{\ensuremath{\tilde{B}}\xspace}
\newcommand{\Zz}{\ensuremath{Z^{0}}\xspace}
\newcommand{\sZino}{\ensuremath{\tilde{Z}^{0}}\xspace}
\newcommand{\sGam}{\ensuremath{\tilde{\gamma}}\xspace}
\newcommand{\chiz}{\ensuremath{\tilde{\chi}^{0}}\xspace}
\newcommand{\chip}{\ensuremath{\tilde{\chi}^{+}}\xspace}
\newcommand{\chim}{\ensuremath{\tilde{\chi}^{-}}\xspace}
\newcommand{\chipm}{\ensuremath{\tilde{\chi}^{\pm}}\xspace}
\newcommand{\Hone}{\ensuremath{H_{d}}\xspace}
\newcommand{\sHone}{\ensuremath{\tilde{H}_{d}}\xspace}
\newcommand{\Htwo}{\ensuremath{H_{u}}\xspace}
\newcommand{\sHtwo}{\ensuremath{\tilde{H}_{u}}\xspace}
\newcommand{\sHig}{\ensuremath{\tilde{H}}\xspace}
\newcommand{\sHa}{\ensuremath{\tilde{H}_{a}}\xspace}
\newcommand{\sHb}{\ensuremath{\tilde{H}_{b}}\xspace}
\newcommand{\sHpm}{\ensuremath{\tilde{H}^{\pm}}\xspace}
\newcommand{\hz}{\ensuremath{h^{0}}\xspace}
\newcommand{\Hz}{\ensuremath{H^{0}}\xspace}
\newcommand{\Az}{\ensuremath{A^{0}}\xspace}
\newcommand{\Hpm}{\ensuremath{H^{\pm}}\xspace}
\newcommand{\sGra}{\ensuremath{\tilde{G}}\xspace}
\newcommand{\mtil}{\ensuremath{\tilde{m}}\xspace}
\newcommand{\rpv}{\ensuremath{\rlap{\kern.2em/}R}\xspace}
\newcommand{\LLE}{\ensuremath{LL\bar{E}}\xspace}
\newcommand{\LQD}{\ensuremath{LQ\bar{D}}\xspace}
\newcommand{\UDD}{\ensuremath{\overline{UDD}}\xspace}
\newcommand{\Lam}{\ensuremath{\lambda}\xspace}
\newcommand{\Lamp}{\ensuremath{\lambda'}\xspace}
\newcommand{\Lampp}{\ensuremath{\lambda''}\xspace}
\newcommand{\spinbd}[2]{\ensuremath{\bar{#1}_{\dot{#2}}}\xspace}

\newcommand{\MD}{\ensuremath{{M_\mathrm{D}}}\xspace}
\newcommand{\Mpl}{\ensuremath{{M_\mathrm{Pl}}}\xspace}
\newcommand{\Rinv} {\ensuremath{{R}^{-1}}\xspace}

%
%
\hyphenation{en-viron-men-tal}

\cmsNoteHeader{09-006}
\title{Time Reconstruction and Performance of the CMS Electromagnetic Calorimeter}

\address[cern]{CERN}

\date{\today}

\abstract{

The resolution and the linearity of time measurements made with the CMS electromagnetic calorimeter are studied with samples of data from test beam electrons, cosmic rays, and beam-produced muons. The resulting time resolution measured by lead tungstate crystals is better than 100~ps for energy deposits larger than 10~GeV. Crystal-to-crystal synchronization with a precision of 500~ps is performed using muons produced with the first LHC beams in 2008.
}

\hypersetup{%
pdfauthor={Seth Cooper, Daniele del Re, Sasha Ledovskoy},%
pdftitle={Time Reconstruction and Performance of the CMS Electromagnetic Calorimeter},%
pdfsubject={CMS},%
pdfkeywords={CMS, ECAL, timing}}

\maketitle 

\section{Introduction}
The primary goal of the Compact Muon Solenoid (CMS) experiment~\cite{CMS} is to explore particle physics at the TeV energy scale, exploiting the proton-proton collisions delivered by the Large Hadron Collider (LHC) at CERN~\cite{LHC}. The electromagnetic calorimeter (ECAL), which measures the energy of electrons and photons produced in LHC collisions, is located inside the bore of the solenoid magnet. It is a hermetic homogeneous calorimeter made of 75\,848 lead tungstate (PbWO$_4$) scintillating crystals:  61\,200 in the barrel (EB) and 7324 in each endcap (EE). The barrel has an inner radius of 129~cm, while the distance between the center of the interaction region and the endcap envelope is about 315~cm. Lead tungstate has a fast scintillation response and is resistant to radiation; it has a high density (8.3~g~cm$^{-3}$), a short radiation length ($X_0$ = 0.89~cm), and a small Moli\`ere radius ($R_M$ = 2.0~cm), features that allow a highly granular, compact detector to be built. Each individual crystal is a truncated pyramid, with a lateral size comparable to $R_M$ and a length of 25.8 $X_0$ (24.7 $X_0$) for the barrel (endcaps). The scintillation decay time of the crystals is comparable to the LHC bunch crossing interval of 25~ns, and about 80\% of the light is emitted in 25~ns. For the light detection, the crystals are equipped with avalanche photodiodes in the barrel and vacuum phototriodes in the endcaps.

The main purpose of the ECAL is the precise energy measurement, needed for many physics analyses. In the barrel region, the target energy resolution for unconverted photons with energies larger than 50~GeV is 0.5\%.  Tests illuminating 25\% of all ECAL barrel crystals with 120~GeV electrons have demonstrated that this target resolution is achievable~\cite{testbeam}. Searches for the Higgs boson particularly benefit from this performance: a Standard Model Higgs with a mass of 120~GeV can be observed by CMS in the two-photon decay channel with a 5$\sigma$ significance with less than 10~fb$^{-1}$ of integrated luminosity collected at 14~TeV center of mass energy~\cite{tdr,tdr2}.

In addition to the energy measurement, the combination of the scintillation timescale of PbWO$_4$, the electronic pulse shaping, and the sampling rate allow excellent time resolution to be obtained with the ECAL. This is important in CMS in many respects. The better the precision of time measurement and synchronization, the larger the rejection of backgrounds with a broad time distribution. Such backgrounds are cosmic rays, beam halo muons, electronic noise, and out-of-time proton-proton interactions. Precise time measurement also makes it possible to identify particles predicted by different models beyond the Standard Model. Slow heavy charged R-hadrons~\cite{hscp}, which travel through the calorimeter and interact before decaying, and photons from the decay of long-lived new particles reach the calorimeter out-of-time with respect to particles travelling at the speed of light from the interaction point. As an example, to identify neutralinos decaying into photons with decay lengths comparable to the ECAL radial size, a time measurement resolution better than 1~ns is necessary. To achieve these goals the time measurement performance both at low energy (1~GeV or less) and high energy (several tens of GeV for showering photons) becomes relevant. In addition, amplitude reconstruction of ECAL energy deposits benefits greatly if all ECAL channels are synchronized within 1~ns~\cite{recoampli}. Previous experiments have shown that it is possible to measure time with electromagnetic calorimeters with a resolution better than 1~ns~\cite{cdftime}.

In Section~\ref{sec:extraction}, the algorithm used to extract the time from the digitized ECAL signal is presented. In Section~\ref{sec:reso}, the uncertainties in the time measurement and the time resolution extracted using electrons from a test beam are detailed. In Section~\ref{sec:synch}, the synchronization of ECAL crystals in preparation for the first LHC collisions is discussed, and the time inter-calibration obtained using muons from the first LHC beam events is presented. Finally, Section~\ref{sec:cosmics} shows results on the ECAL time resolution and linearity, obtained using  cosmic ray muons after the insertion of the ECAL into its final position in CMS.

The scope of this paper is limited to the timing extracted for single crystals. For electromagnetic showers that spread over several crystals, the time measurement can be averaged, thus improving the resolution.
\section{Time extraction with ECAL crystals}
\label{sec:extraction}
\begin{figure}[htbp!]
\centering
\includegraphics[width=0.45\textwidth]
{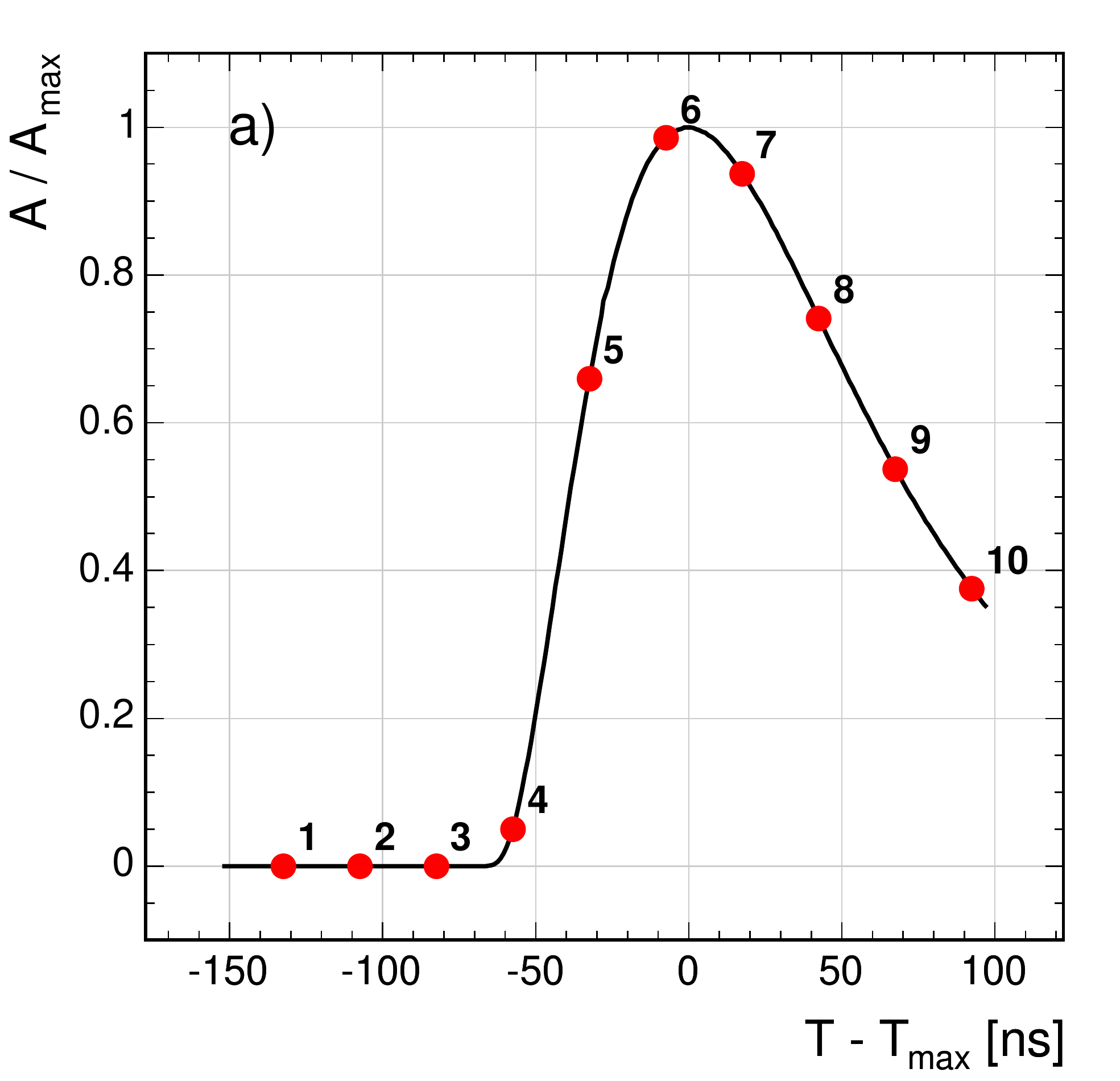}
\includegraphics[width=0.45\textwidth]
{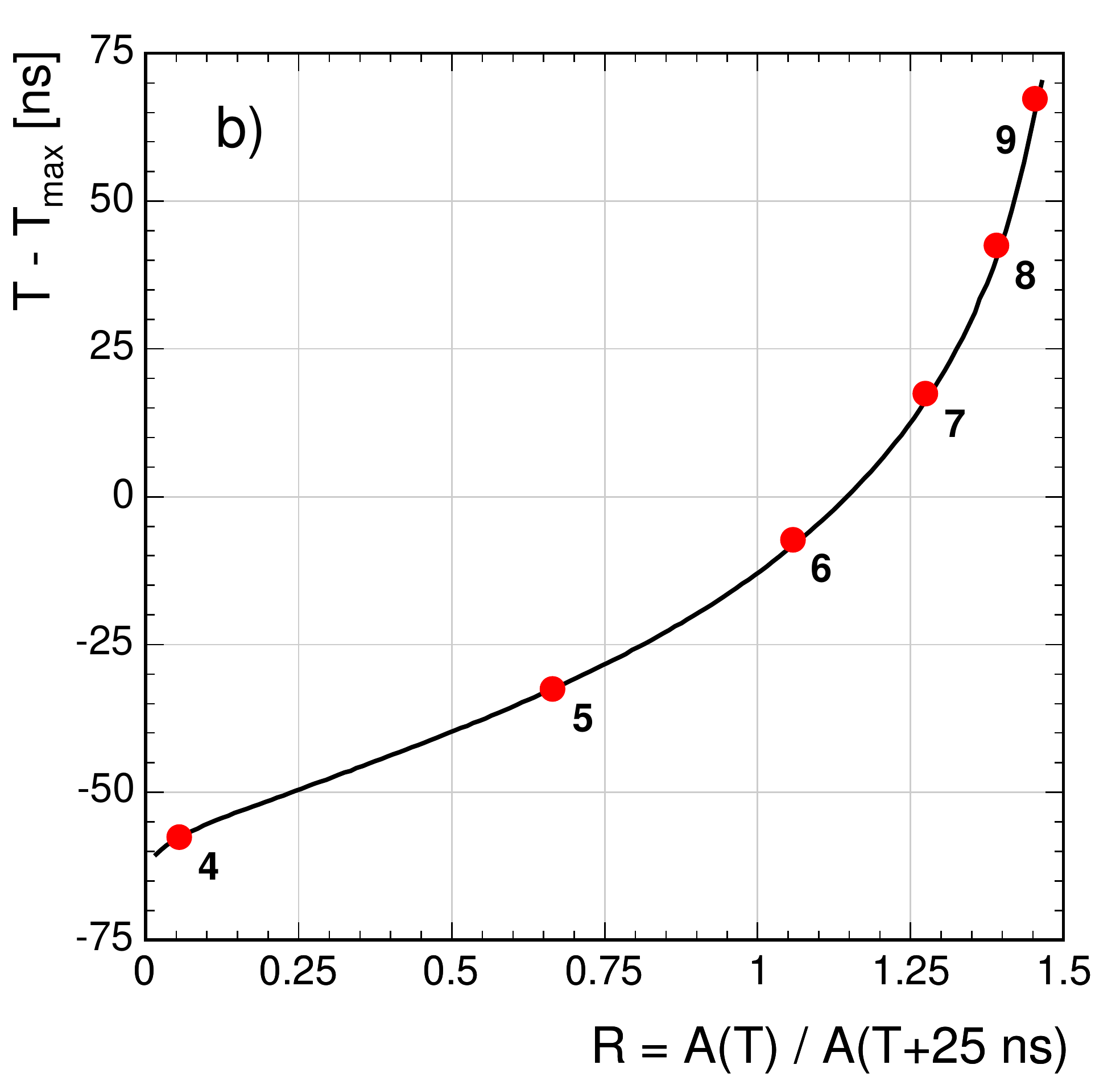}
\caption[Pulse shape]{\label{fig:pulse} (a) Typical pulse shape measured in the ECAL, as a function of the difference between the time ($T$) of the ADC sample and the time ($T_{\rm max}$) of the maximum of the pulse. The dots indicate ten discrete samples of the pulse, from a single event, with pedestal subtracted and normalized to the maximum amplitude. The solid line is the average pulse shape, as measured with a beam of electrons triggered asynchronously with respect to the digitizer clock phase. (b) Pulse shape representation using the time difference $T-T_{\rm max}$ as a function of the ratio of the amplitudes in two consecutive samples ($R$).}
\end{figure}

The front-end electronics of the ECAL amplifies and shapes the signal from the photodetectors~\cite{ecaltdr}. Figure~\ref{fig:pulse}(a) shows the time structure of the signal pulse measured after amplification (solid line). The amplitude of the pulse, $A$, is shown as a function of the time difference $T-T_{\rm max}$, where $T_{\rm max}$ is defined as the time when the pulse reaches its maximum value, $A_{\rm max}$. The pulse shape is defined by the analog part of the front-end electronics. For a given electronic channel, the same pulse shape is obtained, to a very good approximation, for all types of particles and for all momenta. The pulse is then digitized at 40 MHz by a 12-bit voltage-sampling analog-to-digital converter on the front-end, providing a discrete set of amplitude measurements. These samples are stored in a buffer until a Level-1 trigger is received. At that time the ten consecutive samples corresponding to the selected event are transmitted to the off-detector electronics for insertion into the CMS data stream.
In this paper, ECAL time reconstruction is defined as the measurement of $T_{\rm max}$ using the ten available samples of pulse amplitude. For each ECAL channel, the amplitudes of these samples depend on three factors: the value of $A_{\rm max}$; the relative position of $T_{\rm max}$ between time samples, which will be referred to as a ``$T_{\rm max}$ phase"; and the pulse shape itself.

An alternative representation of the pulse shape is provided by a ratio variable, defined as \mbox{$R(T)=A(T)/A(T+25$~ns$)$}. Figure~\ref{fig:pulse}(b) shows the measured pulse shape using the variable $T-T_{\rm max}$, as a function of $R(T)$. In view of the universal character of the pulse shape, this representation is independent of $A_{\rm max}$. It can be described well with a simple polynomial parameterization. The corresponding parameters have been determined in an electron test beam (see Section~\ref{sec:reso}) for a representative set of EB and EE crystals, and are subsequently used for the full ECAL.

Each pair of consecutive samples gives a measurement of the ratio $R_{i}=A_{i}/A_{i+1}$, from which an estimate of $T_{\rm max,i}$ can be extracted, with $T_{\rm max,i} = T_{i} - T(R_{i})$. Here $T_i$ is the time when the sample $i$ was taken and $T(R_{i})$ is the time corresponding to the amplitude ratio $R_i$, as given by the parameterization corresponding to Fig.~\ref{fig:pulse}(b). The uncertainty on each $T_{\rm max,i}$ measurement, $\sigma_i$, is the product of the derivative of the $T(R)$ function and the uncertainty on the value of $R_i$. The latter has three independent contributions, which are added in quadrature.  The first contribution is due to noise fluctuations in each sample. The second contribution is due to the uncertainty on the estimation of the pedestal value subtracted from the measured amplitudes~\cite{recoampli}. The last contribution is due to truncation during 12-bit digitization.

The number of available ratios depends on the absolute timing of a pulse with respect to the trigger. Ratios corresponding to large derivatives of the $T(R)$ function and to very small amplitudes are not used. Pulses from particles arriving in-time with the LHC bunch crossing typically have 4 or 5 available ratios. The time of the pulse maximum, $T_{\rm max}$, and its error are then evaluated from the weighted average of the estimated $T_{\rm max,i}$:
 \begin{equation}
 T_{\rm max} = \frac{\sum_{i} \frac{T_{\rm max,i}}{\sigma_{i}^2}}{
 \sum_{i} \frac{1}{\sigma_{i}^2}} \; \; \; \; \; \; ; \; \;  \; \; \; \;
 \frac{1}{\sigma_{T}^2} = \sum_{i} \frac{1}{\sigma_{i}^2} \quad.
 \label{eq:1}
 \end{equation}
The values of $T_{\rm max,i}$ and their errors $\sigma_i$ are combined as if they were uncorrelated. Adjacent $R_i$ ratios, however, share a common amplitude measurement value, and are thus anti-correlated. Monte Carlo studies show that the uncertainty estimated using Eq.~(\ref{eq:1}) is, on average, about 20\% too large because of the anti-correlation, and that the averaging of individual time measurements 
results in a bias of about 10\% of the statistical uncertainty of $T_{\rm max}$, which is negligible. The different $R_i$ ratios are also correlated because there are correlations in the noise contributions to the samples (see Fig.~3 of Ref.~\cite{recoampli}). This has no impact on the average and a very small effect on the estimated uncertainty of $T_{\rm max}$, corresponding to $<10\%$ of the statistical uncertainty.
\section{Time measurement resolution}
\label{sec:reso}
The time resolution can be expressed as the sum in quadrature of three terms accounting for different sources of uncertainty, and may be parameterized as follows:
\begin{equation}
\sigma^2(t) = \left (\frac{N \sigma_n}{A}\right )^2  + \left (\frac{S}{\sqrt{A}}\right )^2  + C^2 \quad.
\label{eq:reso}
\end{equation}
Here $A$ is the measured amplitude, $\sigma_n$ is related to the noise level in individual samples, and $N$, $S$, and $C$ represent the noise, stochastic, and constant term coefficients, respectively.
The noise term contains the three uncertainties mentioned above, in the discussion of the uncertainty on $T_{\rm max,i}$. Monte Carlo simulation studies give $N=33$~ns, when the electronic noise in the barrel and endcaps is $\sigma_n \sim 42$~MeV and $\sigma_n \sim 140$~MeV, respectively.   The stochastic term comes from fluctuations in photon collection times, associated with the finite time of scintillation emission. It is estimated to be negligible and it is not considered in this study.  The constant term has several contributions: effects correlated with the point of shower initiation within the crystal and systematic effects in the time extraction, such as those due to small differences in pulse shapes for different channels.

To study the pulse shape and determine the intrinsic time resolution of the ECAL detector, electrons from a test beam are used. Several fully equipped barrel and endcap sectors were exposed to electrons at the H2 and H4 test beam facilities at CERN, prior to their installation into the CMS detector~\cite{testbeam}. The beam lines delivered electrons with energies between 15~GeV and 250~GeV. In the test beam, sectors were mounted on a rotating table that allowed the beam to be directed onto each crystal of the supermodule.  The 2-D profile of the electron beam was almost Gaussian, with a spread comparable to the crystal size. As a consequence, in a single run, electrons hit the crystal in different positions and the fraction of energy deposited by an electron in a given crystal varied from event to event.

\begin{figure}[htbp!]
\centering
\includegraphics[width=0.5\textwidth]
{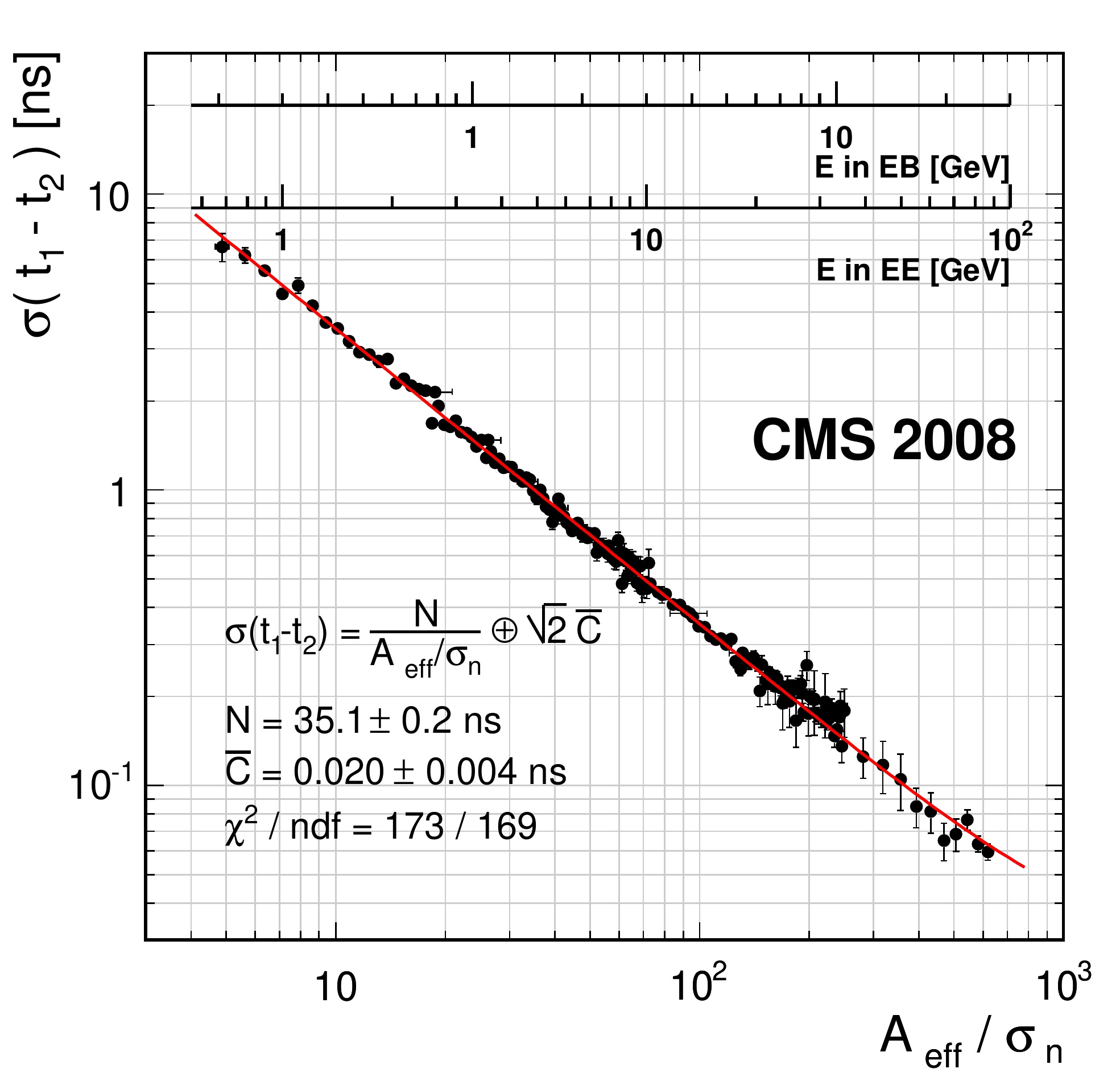}
\caption[]{\label{fig:resolution} Gaussian width of the time difference between two neighboring crystals as a function of the variable $A_{\rm eff}/\sigma_n$, for test beam electrons with energies between 15 and 300~GeV. The equivalent single-crystal energy scales for barrel and endcaps are overlaid on the plot.}
\end{figure}

The time resolution is extracted from the distribution of the time difference between adjacent crystals that share the same electromagnetic shower and measure similar energies. This approach is less sensitive to the constant term $C$, since effects due to synchronization do not affect the spread but only the average of the time difference. As electrons enter the crystal from the front face and there is the requirement of depositing a similar energy in both crystals, the uncertainty due to the variation of the point of shower initiation is also negligible. In addition, the $T-T_{\rm max}$ vs. $R$ polynomial parameterization is determined individually for every crystal to avoid systematic effects due to pulse shape parameterization. 
The distribution of the time difference is well described by a Gaussian 
function with negligible tails for all amplitudes. The spread is 
defined as the sigma of the Gaussian fit to the distribution and 
is parameterized, following Eq.~(\ref{eq:reso}), as
\begin{equation}
 \sigma^2(t_1 - t_2) =  \left (\frac{N \sigma_n}{A_{\rm eff}}\right )^2 + 2 \overline{C}^2
\end{equation}
where $A_{\rm eff}= A_1 A_2 /\sqrt{A_1^2 + A_2^2}$, with $t_{1,2}$ and $A_{1,2}$ corresponding to the times and amplitudes measured in the two crystals, and $\overline{C}$ being the residual contribution from the constant terms.
The extracted width is presented in Fig.~\ref{fig:resolution} as a function of the variable $A_{\rm eff}/\sigma_n$.
The fitted noise term corresponds to $N=(35.1\pm0.2)$~ns. $\overline{C}$ is very small, $\overline{C}=(20\pm4)$~ps. For values of $A_{\rm eff}/\sigma_n$ greater than 400, $\sigma(t)$ is less than 100~ps, demonstrating that, with a carefully calibrated and synchronized detector, it is possible to reach a time resolution better than 100~ps for large energy deposits ($E>$10--20~GeV in the barrel).
As a crosscheck, the stochastic component was left free in the fit and found to be $S<7.9$~ns\;MeV$^{\frac{1}{2}}$ ($90\%$ C.L.), confirming that this term is negligible.
\section{Synchronization between crystals}
\label{sec:synch}
For each individual ECAL channel, the signals generated by particles originating from the interaction point (IP) are registered with approximately the same value of $T_{\rm max}$, because their flight times to the crystal do not change (up to small differences related to the precise position of the IP).  Because the time of flight varies across the ECAL by a few nanoseconds and there are different intrinsic delays among channels, a crystal-to-crystal synchronization of the ECAL must be performed.

The ECAL front-end electronics allows adjustment of $T_{\rm max}$ for groups of 5$\times$5 channels in steps of 1.04~ns.  The determination of values for these adjustments is called hardware synchronization. To take full advantage of the high precision of the ECAL time reconstruction, the value of $T_{\rm max}$ corresponding to particles coming from the IP must be determined for each ECAL channel with an accuracy exceeding the typical time resolution. These additional corrections, called software synchronizations, can be extracted offline with physics collision events. Minimum bias events, which have a typical energy scale of 500 MeV/channel, can be used for this purpose. With the trigger menus planned for early data taking, they will yield about 1000 events/channel/day. A synchronization precision on the order of 100~ps is estimated to be achievable using data from a single day of running at the start of the LHC.

\begin{figure}[htbp!]
\centering
\includegraphics[width=1.\textwidth]
{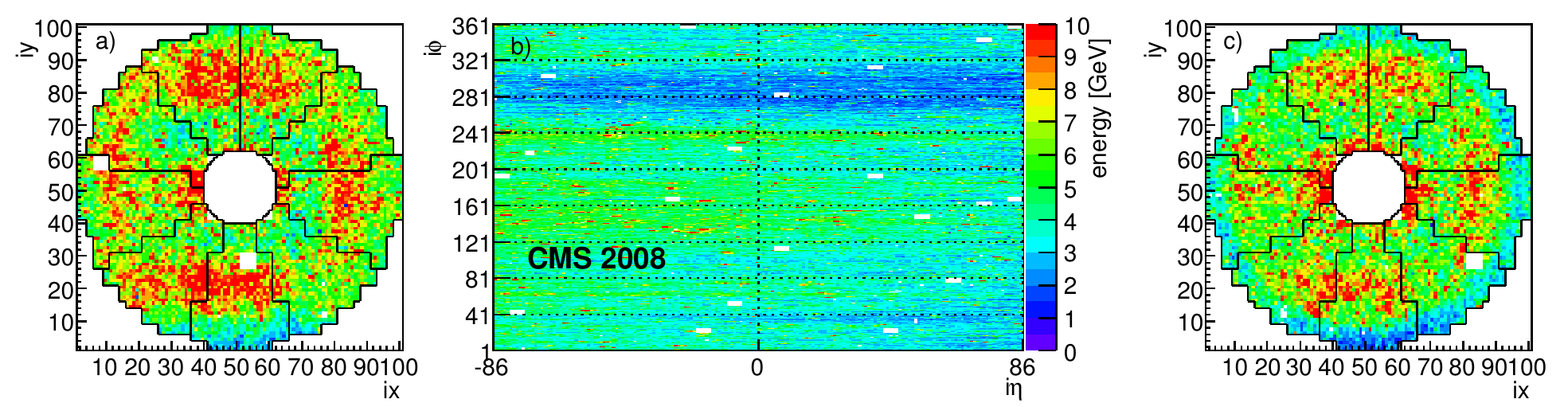}
\caption[Event display]{\label{fig:eventdisplay} ECAL average energy deposit per crystal for a typical ``beam splash'' event with muons coming from the ``minus'' side. (a) Occupancy of the ``minus'' endcap, where $ix$ and $iy$ indicate the indices of the crystals in the horizontal ($x$) and vertical ($y$) coordinates, respectively. (b) Occupancy of the barrel, where  $i\eta$ and $i\phi$ indicate the indices of the crystals in the $\eta$ and $\phi$ coordinates. (c) Occupancy of the ``plus'' endcap. The white regions correspond to channels masked in the readout. They represent a small fraction of the total number of channels, smaller than 1\% in that specific run. Many of these channels have been recovered subsequently.}
\end{figure}

Beam-produced muons, collected by CMS with the first beams circulating in the LHC in September 2008, are used to synchronize the detector. The beams were dumped on collimators located approximately 150 m upstream of CMS, producing so-called ``beam splash'' events. The proton bunch length along the direction of propagation was about 6~cm, corresponding to about 200~ps spread in time. The resulting pions and kaons decayed into a very large number of muons, moving horizontally along the beam direction, corresponding to the $z$ axis, at close to the speed of light. The arrival time of these muons at each crystal depends on the crystal position, and can be precisely predicted.
In Fig.~\ref{fig:eventdisplay} the ECAL energy deposits in each crystal for a typical ``beam splash'' event are shown. Several muons cross each crystal, resulting in energy deposits between 2 and 10~GeV. It may be noted that almost every crystal registered a significant energy.

As stated above, it is important to synchronize the calorimeter such that
particles travelling from the interaction region appear in-time. Since muons from ``beam splash'' events travel as a plane wave and do not come from the interaction region, a correction using the predicted time of flight is applied.  In order to compare times obtained from different events, the average times in the barrel and each endcap are used as references. It should be noted that, because of the time of flight of muons, the ``$T_{\rm max}$ phase'' depends on the position of the crystal and muon direction. Crystals with the same pseudorapidity $\eta$, forming a ring in $\phi$, have a common ``$T_{\rm max}$ phase''.

Two independent samples of ``beam splash'' events are used to synchronize ECAL channels: about 20 events containing a large number of muons travelling in the negative direction of the $z$ axis (``minus" beam, moving clockwise in the LHC) and about 35 events with muons travelling in the opposite direction (``plus" beam). For every individual channel, an average of time measurements weighted by their uncertainties is calculated, resulting in the time intercalibration coefficient. This procedure is applied separately for ``plus" and ``minus" beam events. Comparison of the ``plus" and ``minus" calibrations yields an estimate of the statistical and systematic uncertainties of the calibration and time reconstruction algorithms, while the sum of the two samples is used to extract the intercalibration coefficients.

Figure 4(a) shows the difference between ``plus" and ``minus" calibrations for the 360 barrel channels in which muons arrived at the same time delay with respect to the trigger in both ``plus'' and ``minus'' runs. These channels, forming a ring in $\phi$, have the unique property of sharing the same ``$T_{\rm max}$ phase'' for both ``plus'' and ``minus'' muons. Thus channels in this ring experience conditions similar to those in normal LHC operation i.e.\ the energy deposits are synchronous. The Gaussian spread of the distribution is about 230~ps, which is in good agreement with the expected statistical uncertainty. Summing the event samples from both ``plus" and ``minus" beams results in a synchronization of ECAL channels with a statistical uncertainty of about 85~ps in the barrel and 105~ps in the endcaps.

\begin{figure}[htbp!]
\centering
\includegraphics[width=0.45\textwidth]
{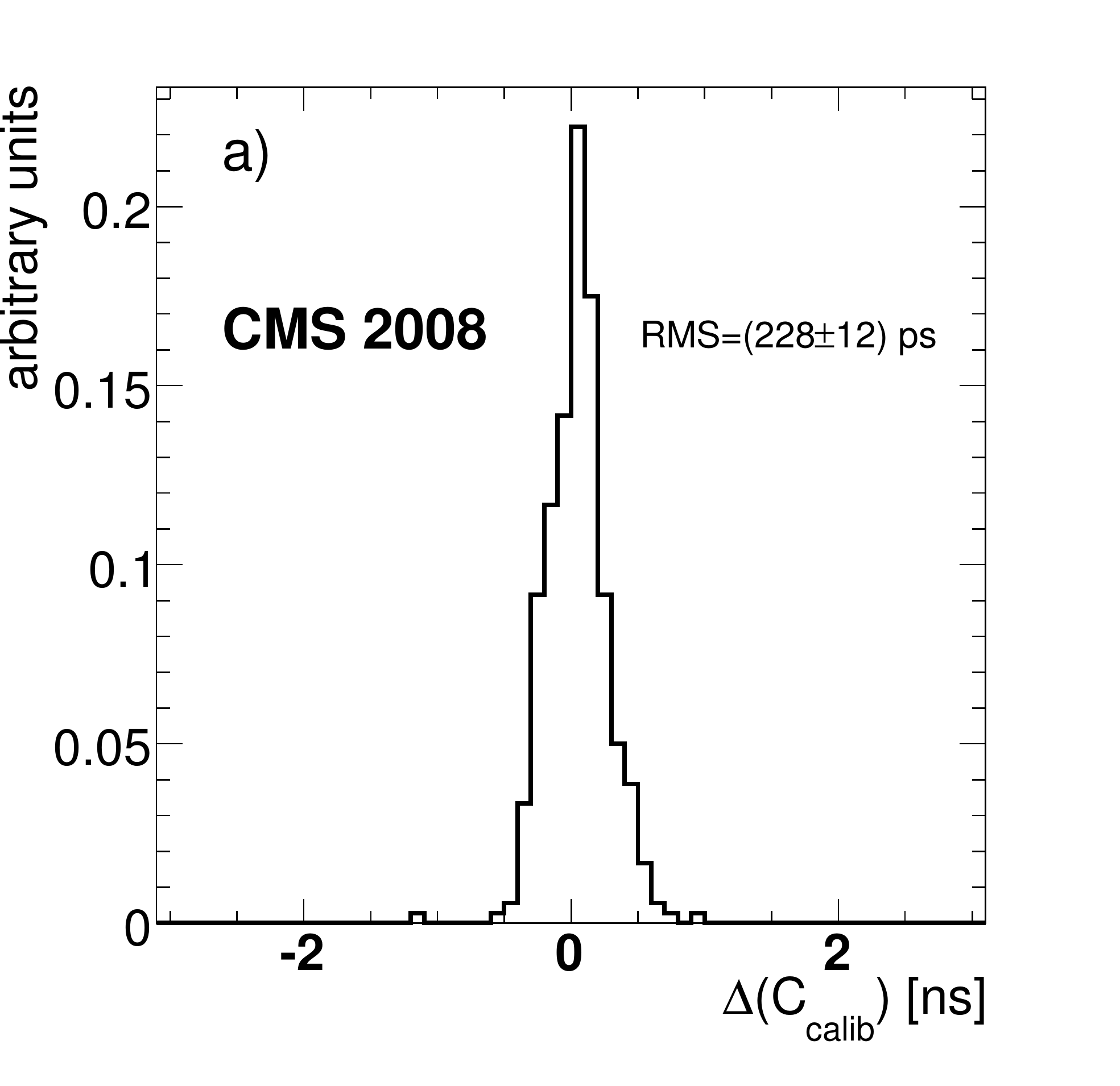}
\includegraphics[width=0.45\textwidth]
{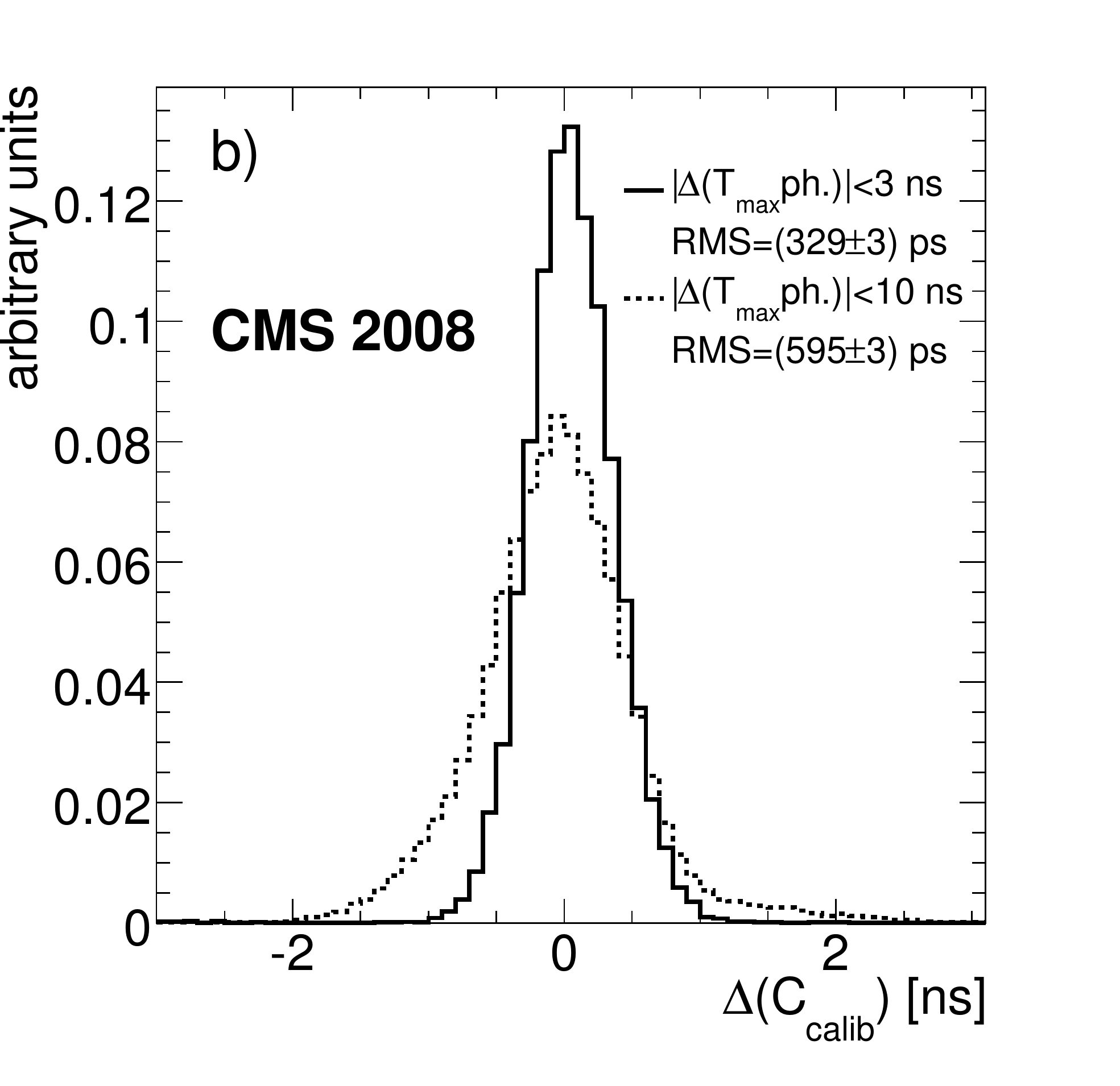}
\caption[]{\label{fig:synch} Distributions of the differences between the calibration coefficients obtained using muons from the ``plus'' beam and muons from the ``minus'' beam for (a) 360 barrel channels in which muons arrived at the same time delay with respect to the trigger, and (b) two different samples of barrel crystals, for which the difference between the mean measured absolute times (t) is in the range 3 to 10~ns (see text). The histograms are normalized to have unit area in each case.}
\end{figure}
Figure~\ref{fig:synch}(b) shows a distribution similar to that in Fig.~\ref{fig:synch}(a), except that muons in these channels need not arrive at the same time in both ``plus" and ``minus" splashes. This has the effect of including many more crystals in the selection and introduces sensitivity to any ``$T_{\rm max}$ phase"-dependent effects. The solid line represents the distribution of channels fulfilling the requirement that the difference in ``$T_{\rm max}$ phase" between ``plus'' and ``minus'' muons is within a 3~ns time range, which includes about 43\% of the barrel channels. The dotted line is the distribution conditioned by requiring a ``$T_{\rm max}$ phase" difference of less than 10~ns, selecting about 70\% of the barrel channels. The widths of these distributions are $(329\pm3)$~ps and $(595\pm3)$~ps, respectively, both of which are significantly larger than the expected statistical uncertainty, indicating the presence of systematic effects correlated with the uncertainties in the pulse shape. The time reconstruction method assumes the same pulse shape for all ECAL channels, but the real pulse shapes slightly differ from channel to channel (see Fig.~10 of Ref.~\cite{recoampli}). Detailed Monte Carlo simulation studies and measurements with electrons from a test beam show that these differences in shape pose no problem for in-time signals, while out-of-time signals are reconstructed with a systematic uncertainty ranging from tens to hundreds of picoseconds. The effect is proportional to the size of the range in ``$T_{\rm max}$ phase". The results shown in Fig.~\ref{fig:synch}(b) confirm these studies. In LHC collisions, the time range will not have a wide spread since events will be synchronous and  the accumulated bias in the time reconstruction will be minimal.  Thus the systematic error on the synchronization is expected to be negligible when using collision events.

It can be concluded that the overall uncertainty in the determination of the synchronization coefficients, which is the quadratic sum of the statistical and systematic uncertainties, is about 300--600~ps. This is the time resolution expected at the start-up of the LHC, when these synchronization coefficients will be used.
\section{Resolution and linearity checks using cosmic ray muons}
\label{sec:cosmics}
The resolution and the linearity of time measurements are determined with a sample of cosmic ray muons collected during summer 2008, when the ECAL was already inserted into its final position in CMS. Samples used for this analysis were taken from runs without magnetic field. Muon tracks are reconstructed in the muon system and, where possible, in the inner tracker. Muons typically deposit energy in several ECAL crystals, which are then grouped to form clusters. The purity of the sample is increased by requiring the extrapolated muon track to point towards the barycenter of the ECAL cluster. This is done by requiring that the distance between the calorimeter deposit and the position of entrance of the muon track in the $\eta$-$\phi$ plane is consistent with zero within the experimental resolution~\cite{muoncruzet}. The selection is restricted to the barrel region, resulting in a sample of about $2 \times 10^5$ muons. The associated clusters correspond to muons that lose energy in the calorimeter by ionization, with very little background contamination. The synchronization constants obtained from ``beam splash'' events are then applied.

The approach to extract the resolution is similar to that described in Section~\ref{sec:reso}, but in this case the crystal with the maximum amplitude is compared with the other crystals in the cluster. Since different pairs of crystals are used, covering the entire barrel, a constant term comparable to the systematic uncertainty of the synchronization is expected.

\begin{figure}[htbp!]
\centering
\includegraphics[width=0.45\textwidth]
{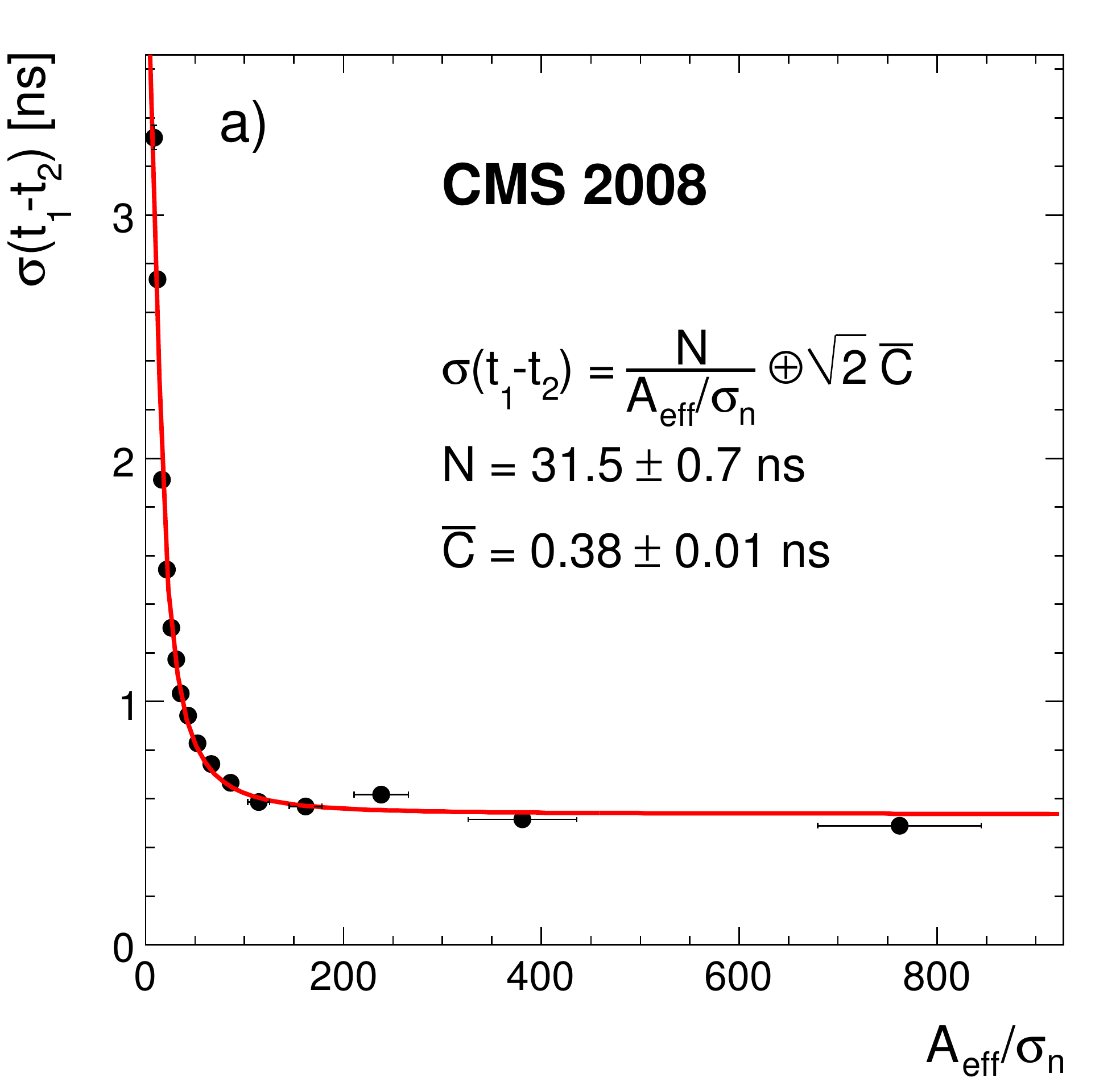}
\includegraphics[width=0.45\textwidth]
{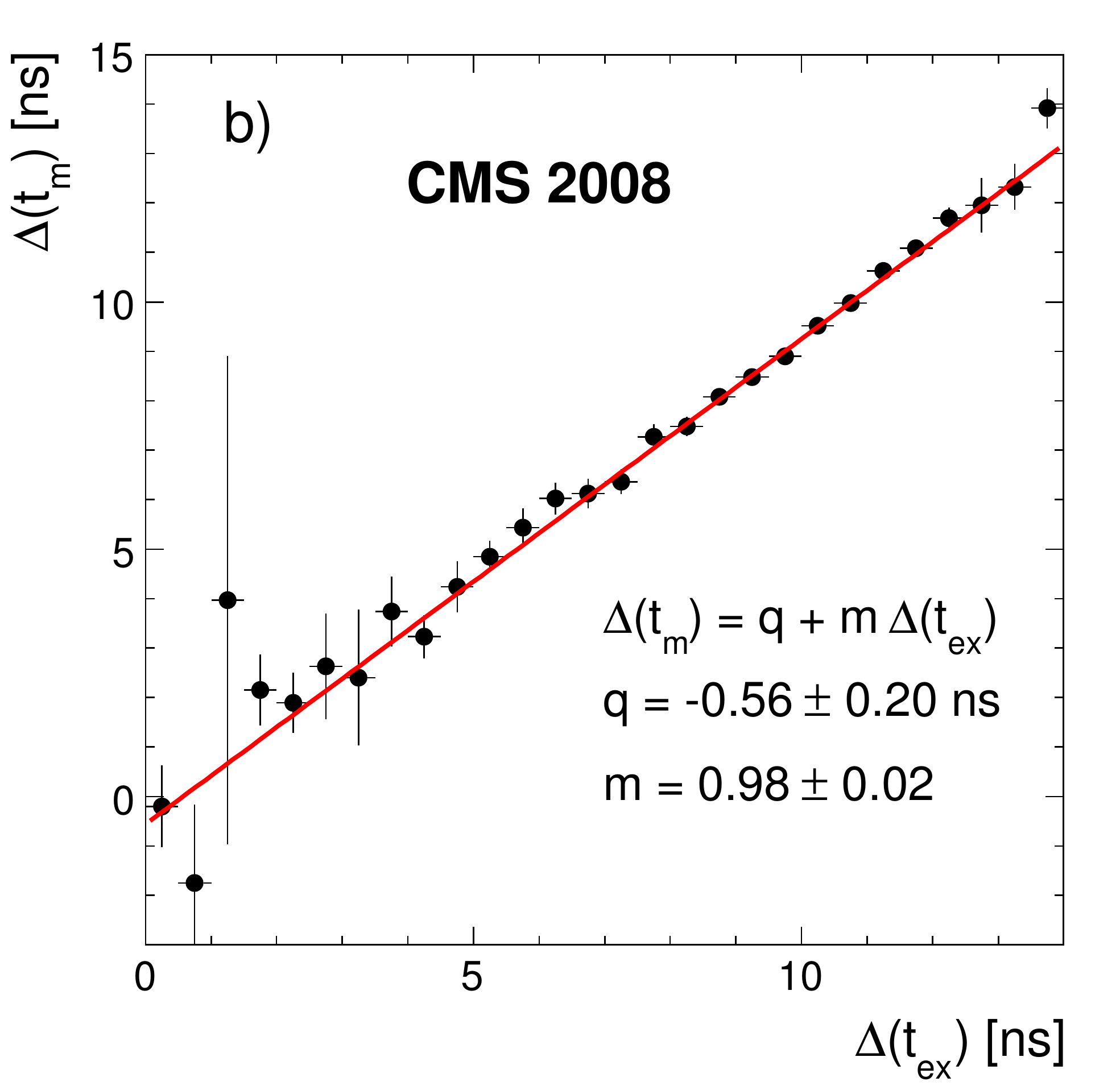}
\caption[]{\label{fig:linearity} (a) Spread of the time difference between crystals of the same cluster, as a function of the variable $A_{\rm eff}= A_1 A_2 /\sqrt{A_1^2 + A_2^2}$, for cosmic ray muons. (b) Measured time difference between top and bottom muon clusters, $\Delta(t_{m})$, as a function of the expected difference corresponding to the time of flight of a relativistic muon, $\Delta(t_{ex})$.}
\end{figure}

The results on the resolution are presented in Fig.~\ref{fig:linearity}(a). The noise term is found to be $N=(31.5\pm0.7)$~ns and is very similar to that obtained from test beam data. The constant term is measured to be $C=(380\pm10)$~ps, which is consistent with the expected systematic uncertainty from ``beam splash'' synchronization.

The same sample of cosmic ray muons is used to test the linearity of the time measurement. For muons which traverse the ECAL barrel from top to bottom, the times of respective clusters are taken to be the times of the crystals with the largest amplitudes. The difference in time between the two crystals is then compared with the corresponding time of flight of a relativistic muon travelling over the distance between the two crystals. The crystals are ordered depending on their vertical position, assuming that all muons are coming from the top of the detector. The distance is calculated taking into account the fact that, on average, cosmic ray muons enter crystals at the center of the lateral edge. The time of flight ranges from about 0~ns, which corresponds to muons almost tangential to the ECAL surface, to  about 14~ns. In Fig.~\ref{fig:linearity}(b) the correlation between expected and measured times is shown. The distribution is fitted with a straight line, resulting in a slope ($m$) compatible with unity. The offset ($q$) is compatible with zero within the systematic uncertainty on the synchronization, which is of the order of 300--600~ps, as discussed in Section~\ref{sec:synch}.
\section{Conclusions}
The resolution and the linearity of the time measurement of the CMS electromagnetic calorimeter have been investigated with samples of data from test beam electrons, cosmic rays, and ``beam splash'' events.
Results obtained with test beam electrons show that the resolution for electromagnetic showers, which can be reached with a perfect time alignment, is better than 100~ps for large energies (more than 10--20~GeV in the barrel). At lower energies, the noise term limits the resolution. As an example, 1~GeV energy deposits in the ECAL barrel have a time resolution of 1.5~ns. The noise term measurement has been confirmed using cosmic ray muon events with the ECAL detector fully equipped and inserted in CMS.  The linearity of the time measurement has been verified using cosmic ray muons that travel across the ECAL barrel, by comparing the measured time difference between the top and the bottom parts of the detector with the expected muon time of flight.

``Beam splash'' events have been used to synchronize all ECAL crystals with a precision of $\sim$500~ps. The corresponding set of synchronization coefficients will be used at LHC start-up. The synchronization will be much improved once collision data are available.

In summary, in addition to measuring the energy of electromagnetic particles with high resolution, the CMS ECAL also provides precise timing information, which will be important for additional background rejection and discoveries of new physics with time-sensitive signatures.
\section*{Acknowledgments}
We thank the technical and administrative staff at CERN and other CMS Institutes, and acknowledge support from: FMSR (Austria); FNRS and FWO (Belgium); CNPq, CAPES, FAPERJ, and FAPESP (Brazil); MES (Bulgaria); CERN; CAS, MoST, and NSFC (China); COLCIENCIAS (Colombia); MSES (Croatia); RPF (Cyprus); Academy of Sciences and NICPB (Estonia); Academy of Finland, ME, and HIP (Finland); CEA and CNRS/IN2P3 (France); BMBF, DFG, and HGF (Germany); GSRT (Greece); OTKA and NKTH (Hungary); DAE and DST (India); IPM (Iran); SFI (Ireland); INFN (Italy); NRF (Korea); LAS (Lithuania); CINVESTAV, CONACYT, SEP, and UASLP-FAI (Mexico); PAEC (Pakistan); SCSR (Poland); FCT (Portugal); JINR (Armenia, Belarus, Georgia, Ukraine, Uzbekistan); MST and MAE (Russia); MSTDS (Serbia); MICINN and CPAN (Spain); Swiss Funding Agencies (Switzerland); NSC (Taipei); TUBITAK and TAEK (Turkey); STFC (United Kingdom); DOE and NSF (USA). Individuals have received support from the Marie-Curie IEF program (European Union); the Leventis Foundation; the A. P. Sloan Foundation; and the Alexander von Humboldt Foundation.
\bibliography{auto_generated}   

\providecommand{\href}[2]{#2}\begingroup\raggedright\begin{thebibliography}{10}

\bibitem{CMS}
{CMS collaboration}, ``{The CMS experiment at the CERN LHC}'', {\em JINST} {\bf
  3} (2008) S08004. \href{http://dx.doi.org/10.1088/1748-0221/3/08/S08004}{{\tt
  doi:10.1088/1748-0221/3/08/S08004}}.

\bibitem{LHC}
L.~Evans, (ed.~) and P.~Bryant, (ed.~), ``{LHC Machine}'', {\em JINST} {\bf 3}
  (2008) S08001. \href{http://dx.doi.org/10.1088/1748-0221/3/08/S08001}{{\tt
  doi:10.1088/1748-0221/3/08/S08001}}.

\bibitem{testbeam}
P.~Adzic {et~al.}, ``{Energy resolution of the barrel of the CMS
  electromagnetic calorimeter}'', {\em JINST} {\bf 2} (2007) P04004.
  \href{http://dx.doi.org/10.1088/1748-0221/2/04/P04004}{{\tt
  doi:10.1088/1748-0221/2/04/P04004}}.

\bibitem{tdr}
{CMS collaboration}, ``{CMS Technical Design Report, Volume I: Detector
  Performance and Software}'', {\em CERN-LHCC-2006-001} (2006).

\bibitem{tdr2}
{CMS collaboration}, ``{CMS Technical Design Report, Volume II: Physics
  Performance}'', {\em J. Phys.} {\bf G34} (2007) 995--1579.
  \href{http://dx.doi.org/10.1088/0954-3899/34/6/S01}{{\tt
  doi:10.1088/0954-3899/34/6/S01}}.

\bibitem{hscp}
R.~Mackeprang and A.~Rizzi, ``{Interactions of coloured heavy stable particles
  in matter}'', {\em Eur. Phys. J.} {\bf C50} (2007) 353--362.

\bibitem{recoampli}
P.~Adzic {et~al.}, ``{Reconstruction of the signal amplitude of the CMS
  electromagnetic calorimeter}'', {\em Eur. Phys. J.} {\bf C46S1} (2006)
  23--35. \href{http://dx.doi.org/10.1140/epjcd/s2006-02-002-x}{{\tt
  doi:10.1140/epjcd/s2006-02-002-x}}.

\bibitem{cdftime}
M.~Goncharov {et~al.}, ``{The Timing system for the CDF electromagnetic
  calorimeters}'', {\em Nucl. Instrum. Meth.} {\bf A565} (2006) 543--550.

\bibitem{ecaltdr}
{CMS collaboration}, ``{The Electromagnetic Calorimeter Project: Technical
  Design Report}'', {\em CERN-LHCC-97-33} (1997).

\bibitem{muoncruzet}
{CMS collaboration}, ``{Measurement of the muon stopping power in Lead
  Tungstate}'', {\em submitted to JINST}.

\end{thebibliography}\endgroup

\cleardoublepage\appendix\section{The CMS Collaboration \label{app:collab}}\begin{sloppypar}\hyphenpenalty=500\textbf{Yerevan Physics Institute,  Yerevan,  Armenia}\\*[0pt]
S.~Chatrchyan, V.~Khachatryan, A.M.~Sirunyan
\vskip\cmsinstskip
\textbf{Institut f\"{u}r Hochenergiephysik der OeAW,  Wien,  Austria}\\*[0pt]
W.~Adam, B.~Arnold, H.~Bergauer, T.~Bergauer, M.~Dragicevic, M.~Eichberger, J.~Er\"{o}
% ...
\vskip\cmsinstskip
% ...

\end{sloppypar}
\end{document}